\documentclass[english,reprint,aps,pra,superscriptaddress,longbibliography]{revtex4-2}

\usepackage{graphicx,mathrsfs,enumerate,epstopdf}
\usepackage[utf8]{inputenc}
\usepackage{mathtools}
\usepackage{amsfonts,amstext}
\usepackage{amsmath,amsthm,amssymb,tensor,diagbox,multirow}
\usepackage[hidelinks]{hyperref}\hypersetup{pdfpagemode=UseNone,pdfstartview=}
\usepackage[left=2cm,right=2cm]{geometry}
\usepackage{xcolor}
\usepackage{url}
\usepackage{times}
\usepackage{soul}
\usepackage{dsfont}

\makeatletter

\DeclareMathOperator{\Tr}{Tr}

\makeatother

\usepackage{babel}

\begin{document}
\title{Local quantum overlapping tomography}
\author{Bruna G. M. Araújo}
\thanks{These authors contributed equally to this work.}
\affiliation{ICFO - Institut de Ciencies Fotoniques, The Barcelona Institute of Science and Technology, 08860, Castelldefels, Barcelona, Spain}
\affiliation{Department of Physics, Universitat Aut\`onoma de Barcelona, Campus UAB, Bellaterra, 08193 Barcelona, Spain}

\author{Márcio M. Taddei}
\thanks{These authors contributed equally to this work.}
\affiliation{ICFO - Institut de Ciencies Fotoniques, The Barcelona Institute of Science and Technology, 08860, Castelldefels, Barcelona, Spain}
\email{marciotaddei [at] gmail [dot] com}

\author{Daniel Cavalcanti}
\affiliation{ICFO - Institut de Ciencies Fotoniques, The Barcelona Institute of Science and Technology, 08860, Castelldefels, Barcelona, Spain}
\affiliation{Bitflow, Piquer 23, Barcelona, 08004 Spain}
\affiliation{Algorithmiq Ltd, Kanavakatu 3 C, FI-00160 Helsinki, Finland}

\author{Antonio Acín}
\affiliation{ICFO - Institut de Ciencies Fotoniques, The Barcelona Institute of Science and Technology, 08860, Castelldefels, Barcelona, Spain}
\affiliation{ICREA - Instituci\'o Catalana de Recerca i Estudis Avan\c cats, Lluis Companys 23, 08010 Barcelona, Spain}

\date{\today}
\begin{abstract}
Reconstructing the full quantum state of a many-body system requires the estimation of a number of parameters that grows exponentially with system size. Nevertheless, there are situations in which one is only interested in a subset of these parameters and a full reconstruction is not needed. A paradigmatic example is a scenario where one aims at determining all the reduced states only up to a given size. Overlapping tomography provides constructions to address this problem with a number of product measurements much smaller than what is obtained when performing independent tomography of each reduced state. There are however many relevant physical systems with a natural notion of locality where one is mostly interested in the reduced states of neighboring particles. In this work, we study this form of local overlapping tomography. First of all, we show that, contrary to its full version, the number of product-measurement settings needed for local overlapping tomography does not grow with system size. Then, we present strategies for qubit and fermionic systems in selected lattice geometries. The developed methods find a natural application in the estimation of many-body systems prepared in current quantum simulators or quantum computing devices, where interactions are often local.
\end{abstract}
\maketitle


\section{Introduction}
\label{sec:intro}

There have been great advances in the construction of quantum devices, with systems composed of tens of individually addressable qubits \cite{Arute2019,Kelly2019,Arute2020,Wu2021,Cho2020}, 
more than a hundred bosonic modes \cite{Zhong2020,Zhong2021}, or intermediate-scale systems of up to thousands of entangled qubits \cite{Britton2012,DWave2000Q}, with further advances clearly on the way. These are based on several different physical systems, such as trapped-ion spins, superconductors or photons. 
These developments bring to the fore the problem of efficiently measuring a quantum system. A full description of a system of size $n$ requires an amount of parameters exponential in $n$, and likewise full quantum state tomography requires an exponential amount of resources (e.g.\ measurement settings, outcomes), a feat that has been achieved so far for a maximum of 10 qubits \cite{Song2017}.

Several approaches have been presented to circumvent the problem and reconstruct the quantum state with fewer measurements settings and fewer produced copies of the state to be measured. 
Some assume an \emph{a priori} structure of the measured states, such as classes including matrix-product states \cite{Cramer2010,Lanyon2017}, or require certain structures for efficient measurement \cite{Torlai2018,Carrasquilla2019,Smart2021}. 
The ``shadow tomography'' technique \cite{Aaronson2018,Brandao2019} is not at its most efficient when reconstructing quantum states, and additionally requires a quantum memory to store several copies of the state and act collectively on them, or else its advantage is lost \cite{Chen2021}.  Randomized techniques \cite{Evans2019,Huang2020} aim at obtaining expectation values of given operators under a quantum state, but can be used for state reconstruction as well.

A less ambitious, but possibly more realistic approach is to focus not on the entire quantum state but on some of its properties of interest. A natural example is that where one is interested in estimating its $k$-body reduced density matrices ($k$-RDMs). It is not uncommon for there to be interest in $k$-body operators, e.g.\ to detect the entanglement~\cite{Toth2007,Toth2009} or Bell non-locality~\cite{Tura2014} of the many-body state from the expectation values of $2$-body operators. A direct approach consists of measuring each $k$-RDM independently. If we restrict our considerations to products of single-particle projective measurements for their experimental feasibility, this already produces a significant reduction of measurement settings compared to full tomography: the latter requires $e^{\mathcal O(n)}$ measurement settings, while each $k$-RDM requires $e^{\mathcal O(k)}$ measurement settings, hence the complete set of $k$-RDMs, $e^{\mathcal O(k)}\binom nk \sim e^{\mathcal O(k)}(n^k/k^k)$ settings. 
However, existing techniques \cite{Yu2019,Bonet-Monroig2020,Garcia-Perez2020,Cotler2020,Evans2019,Tilly2021,Jiang2020,Zhao2020} based on parallelization can reduce that amount substantially. For a system of $n$ qubits, the technique in \cite{Bonet-Monroig2020} uses at most $\mathcal O(e^k \log^{k-1} n)$ measurement settings to obtain all $k$-RDMs, and the one in \cite{Garcia-Perez2020} obtains a constant-factor reduction of the latter for $k=2$; in \cite{Cotler2020} it is shown that $e^{\mathcal O(k)}\log n$ measurement settings suffice. 
For systems of fermions, the $\binom{2n}{2k}$ operators needed can be measured \cite{Bonet-Monroig2020,Tilly2021} with $\mathcal O(n^2)$ measurement settings if $k=2$, and for any $k$ there are schemes \cite{Jiang2020} with $\mathcal O(n^k)$ or, even better, $\mathcal O(\binom nk)$ measurement settings \cite{Zhao2020}. It is noticeable that restrictions to parallelism from fermionic anticommutation relations negatively impact the reduction of complexity. 
To the best of our understanding, none of these results is known to be tight.
 
The measurement strategies above target obtaining every $k$-RDM of an $n$-partite system. However, there are often cases where certain subsets are especially relevant and others, much less so. The most typical example is when there is a spatial distribution (e.g.\ a lattice) with local interactions expected (though not necessarily assumed) to play an important role: RDMs containing neighboring subsystems are much more important than those containing faraway ones. The main goal of this work is to provide efficient strategies to measure these reduced states by means of products of local measurements, as in overlapping tomography. Our first result is to show that this form of \emph{local quantum overlapping tomography} (local QOT) produces a much more dramatic reduction in complexity: while the naive approach in which full tomography is performed for all reduced states of interest requires a number of measurement settings that grows with the number of reduced states, and hence with the system size, an efficient parallelization can do away with the system-size dependence. After presenting this general result, we provide finer measurement strategies for finite-dimensional and fermionic systems in well-known lattice geometries.

It is propitious at this point to mention scenarios where local overlapping tomography is more (and less) promising as a measurement tool. Contexts in which long-range correlations are expected are not well-suited for direct use of local QOT, since it deliberately forgoes probing them. Nevertheless, in the case of constructed quantum systems (i.e.\ quantum circuits), individual $k$-body gates may be characterized with much parallelization by (few rounds of) local QOT.

 The article is structured as follows:  In Sec.\ \ref{sec:prelim}, we present existing results and set the notation. In Sec.\ \ref{sec:main} we present our main results, first in the form of a general tiling argument, then for qubits (Sec.\ \ref{sec:qubits}) and lastly for fermions (Sec.\ \ref{sec:fermions}). In Sec.\ \ref{sec:conclusion} we make our concluding remarks.

\section{Preliminaries}
\label{sec:prelim}

Our main goal is to find descriptions of $k$-body subsystems with as few measurement settings as possible. In an experimental scenario, the amount of different measurement settings typically impacts the overall experimental complexity, and its reduction may turn experimental proposals from potential to actually feasible. Because we want to obtain practical schemes with experimental feasibility, we only consider products of projective local measurements. This excludes entangling measurements, notably harder to perform, as well as constructions defined by SIC-POVMs (symmetric, informationally complete positive operator-valued measures), which reduce the number of different measurements at the cost of a markedly higher complexity of the experimental setup ~\cite{Durt2008,Medendorp2011,Hou2018}. As such, one measurement setting for the system consists in a specific sequence of single-particle measurement settings. 
Let us first review some existing results, and with it fix some relevant definitions and notation. In what follows, when considering finite-dimensional systems, we restrict our considerations to qubits, although most of our results can easily be generalized to arbitrary dimension. 

\subsection{Qubits}
\label{sec:prelimqubits}
 If $\rho$ is an $n$-qubit state, any state obtained by taking the partial trace with respect to a set of $n-k$ qubits is a $k$-RDM of $\rho$. It is possible to reconstruct such $k$-RDMs from the expectation values of products of Pauli matrices. Consider as an example the $k$-RDM of the first $k$ qubits, $\Tr_{[n]\backslash[k]}(\rho)$, where we define $[j]:=\{0,1,\cdots,j-1\}$. The set of expectation values of the form $\langle\sigma_{a_0}^{(0)}\otimes\cdots\otimes\sigma_{a_j}^{(j)}\otimes\cdots\otimes\sigma_{a_{k-1}}^{(k-1)}\rangle$ clearly suffices to characterize this $k$-RDM if $\sigma_{a_j}^{(j)}$ can be either the identity or one of the Pauli matrices $\sigma_x$, $\sigma_y$, $\sigma_z$ of the $j$-th qubit. 
However, this can be simplified by the fact that the algorithms in question are based on making local measurements with access to each local outcome. In this scenario, any measurement setting that includes an identity can be simulated by taking the suitable marginal of a setting with a Pauli matrix in its place. Thus, obtaining the $k$-RDM $\Tr_{[n]\backslash[k]}(\rho)$ is equivalent to obtaining the expectation values in the form $\sigma_{a_0}^{(0)}\otimes\cdots\otimes\sigma_{a_j}^{(j)}\otimes\cdots\otimes\sigma_{a_{k-1}}^{(k-1)}$, with $\sigma_{a_j}^{(j)}$ being only the three Pauli matrices $\sigma_x$, $\sigma_y$, $\sigma_z$ of the $j$-th qubit.

For a system of $n$ qubits, several works have tackled the issue of obtaining all $\binom nk$ existing $k$-RDMs with as few different measurement settings as possible through parallelization. For the $2$-RDM, Refs.\ \cite{Cotler2020,Bonet-Monroig2020,Garcia-Perez2020} have obtained $\langle\sigma_a^{(i)}\otimes\sigma_b^{(j)}\rangle$ for all pairs of qubits $(i,j)$ and $a,b$ covering all combinations of $\sigma_x$, $\sigma_y$, $\sigma_z$. Refs.\ \cite{Bonet-Monroig2020} and \cite{Cotler2020} present schemes with $N=3+6\lceil\log_2n\rceil$ measurement settings based on partitioning and a family of hash functions $(n,2)$, respectively; \cite{Garcia-Perez2020} reduces that scaling to $N=3+6\lceil\log_3n\rceil$. The first two allow for generalizations to $k>2$: \cite{Bonet-Monroig2020} presents a scheme based on recursive partitioning with $N=\mathcal O(3^k\log_2^{k-1}n)$ measurement settings; \cite{Cotler2020} reduces the problem to finding hash functions $(n,k)$, yielding $N=e^{\mathcal O(k)}\log_2n$ measurement settings (see Section \ref{sec:numrepetitions} for a further result). 

All these results target the obtention of all $k$-RDMs, regardless of any structure. As mentioned, we follow here a different approach, where the RDMs of interest are those of neighboring sites on a lattice. Many relevant systems, such as ion traps \cite{Britton2012,Bruzewicz2019}, quantum-dot lattices \cite{Hensgens2017}, semiconductor lattices \cite{Singha2011}, among others, work based on local interactions with neighbors (first or beyond), and would benefit from a measurement scheme focused on them.

\subsection{Fermions}
\label{sec:prelimfermions}
In addition to qubits, we also tackle the efficient measurement of systems composed of fermions. Fermionic systems present a much greater challenge than qubits. Because of their anticommutation relations, operators on different lattice sites may no longer commute, hence one cannot rely on tensor-product relations to ensure jointly measurability, as in qubits.

In a fermionic system of $n$ modes with $m$ particles ($n\geqslant m$), the $k$-RDM is given by tracing out $(m-k)$ particles, $\Tr_{[m-k]}(\rho)$, with the difference that any set of $(m-k)$ particles is equivalent, due to symmetry. Each mode $i$ has an annihilation ($a_i$) and a creation ($a_i^\dagger$) operator, and to obtain $\Tr_{[m-k]}(\rho)$ it suffices~\cite{Rubin2018} to obtain the expectation values of operators composed of $k$ creation and $k$ annihilation operators. For the 1-RDM, for instance, the values of $\langle a_i^\dagger a_j\rangle$ for all modes $i,j$ suffice. It will be convenient, however, to work with Majorana operators \cite{Bravyi2002,Vidal2003,Kraus2009}, which offer an equivalent description of fermionic systems,
\begin{equation}
\gamma_{2j} := a_j^\dagger+a_j \ , \quad \quad \gamma_{2j+1}:=i(a_j^\dagger-a_j) \ .
\label{eq:defMajorana}
\end{equation}
Majorana operators (Majoranas, for short) are Hermitian with $\pm1$ eigenvalues and have an important role in the (partial) equivalence between fermions and qubits \cite{Jordan1928,Bravyi2002,Verstraete2005}. The $k$-RDM of a given system requires all expectation values $\langle\gamma_p\gamma_q\cdots\gamma_r\gamma_s\rangle$ composed of $2k$ Majoranas, e.g.\ $\langle\gamma_p\gamma_q\rangle$ for all $p,q$ for the 1-RDM. Of utmost relevance for parallelization is that different Majoranas all anticommute, a marked difference to qubits.

Once again, the existing literature tackles the obtention of all $\langle\gamma_p\gamma_q\cdots\gamma_r\gamma_s\rangle$ for a given value $k$. By appropriately pairing Majoranas, Ref.~\cite{Bonet-Monroig2020} obtains so-called ``cliques'' composed of commuting multiple-Majorana operators. As all operators in a clique are Hermitian and commute, they can be determined in a single measurement setting, hence only one setting is required per clique. There are established routines to perform such measurements, typically with fermion-to-qubit mappings~\cite{Jiang2020,Zhao2020}.

For the 1-RDM, the authors of~\cite{Bonet-Monroig2020} present a scheme to cover all needed Majorana operators with $2n-1$ commuting cliques ($N=2n-1$) and for the 2-RDM, with $(10/3)n^2+\mathcal O(n)$ commuting cliques. A general lower bound of $N=\binom{2n}{2k}/\binom{n}{k}$ on the needed number of cliques (hence, of measurement settings) is also shown in \cite{Bonet-Monroig2020}; for the 1-RDM it is saturated by their scheme, $\binom{2n}{2}/\binom{n}{1}=2n-1$, whereas for the 2-RDM the bound is below their scheme by a prefactor in the leading-order term, $\binom{2n}{4}/\binom{n}{2}=(4/3)n^2+\mathcal O(n)$. 

In contrast, here we consider fermions in a lattice~\cite{Imada1998,Jordens2008,Singha2011,Hensgens2017,Schuch2019}; to each lattice site $j$ belong two Majorana operators, as in Eq.\ \eqref{eq:defMajorana}. Given the spatial structure of the lattice, we tackle measurements of neighbors only, e.g.\ terms such as $\langle\gamma_i\gamma_j\rangle$ where $\gamma_i$, $\gamma_j$ belong to neighbor sites. Even though this is not sufficient to fully obtain a $k$-RDM, there is a great interest in efficiently obtaining such expectation values, which are relevant to survey quantities that hinge on groups of close neighbors, like coupling energies. Naturally, by focusing on a smaller subset of modes, the scaling of the number of measurement settings drastically reduces.

\subsection{On the total number of repetitions}
\label{sec:numrepetitions}

The results summarized above concern the count $N$ of measurement settings, e.g.\ the count of on how many different Pauli bases one must measure. It is also relevant to consider the number of repetitions $M$ of each setting needed for accurate statistics. The product $NM$ is the total number of measurement rounds, or equivalently, the total number of state copies needed. Using the Chernoff-Hoefding bound, the reasoning from \cite{Cotler2020} can show that $M\sim\frac2{\varepsilon^2}[k\log n+\log(1/\delta)]$ repetitions suffice to measure a full set of $k$-RDMs such that, with probability at least $1-\delta$, the error in every local measurement is at most $\varepsilon$. Adapting that calculation to our localized $k$-RDMs, we see that the number of repetitions $M$ can be set at $M\sim \frac2{\varepsilon^2}[k+\log n+\log(1/\delta)]$ (see Appendix \ref{sec:ChernoffHoeffding4us}).

In light of this distinction, the results in \cite[Theorem 3]{Evans2019} and in \cite{Huang2020} can be properly appreciated. Applied to a $k$-RDM, the former states that the total number of rounds $NM$ scales as $NM\sim\frac2{\varepsilon^2}3^k\log(3^k\binom nk)\sim\frac2{\varepsilon^2}\mathcal O(3^k)\log n$, lower than the literature results mentioned above, for which $NM\sim\log^2n$ or higher. The method of ``classical shadows'' of the latter can also be applied to $k$-RDMs and similarly yields the scaling of $NM = \mathcal O(\frac1{\varepsilon^2}3^k\log n)$.

The need for $M$ repetitions is an extra reason to determine the paradigm of projective local measurements. For the sake of the argument, consider that three measurement settings on a qubit, one on each Pauli basis, are described as a single POVM with probability $\frac13$ for each basis. Allowing such POVM would reduce the number of measurement settings $N$ on a qubit to a third, but would evidently triple the amount of repetitions $M$, leaving the product $N M$ unchanged. Forbidding nonprojective measurements prevents such artificial reduction of $N$. Entangling measurements and SIC-POVMs reduce $N$ in a way that is not as artificial~\cite{Jiang2020,Garcia-Perez2021}, but irrespective of how $M$ is affected, these are ruled out for their much more complex practical implementation. 
For the remainder of this paper, we will discuss the number $N$ of measurement settings, with the number of repetitions $M$ implicitly assumed. 
 
\section{Local Overlapping Tomography}
\label{sec:main}

As mentioned, we are interested in many-particle systems with a natural notion of locality or distance. The standard way of encapsulating this notion is through lattices, where each particle is located in a node of the lattice.
Therefore, the general question of efficient local overlapping tomography is to, out of an $n$-site lattice, obtain the RDMs of all sets of $k$ neighboring sites arranged in a given shape ($k$-sets). As shown in previous results, full $k$-RDMs require a number $N$ of measurement settings that grows with system size $n$. However, when the target are the $k$-RDMs of neighboring lattice sites, there are efficient parallelization schemes that significantly reduce $N$ to the point of removing the $n$-dependence altogether. Let us see how this can happen.

\begin{figure}
\includegraphics[width=.75\columnwidth]{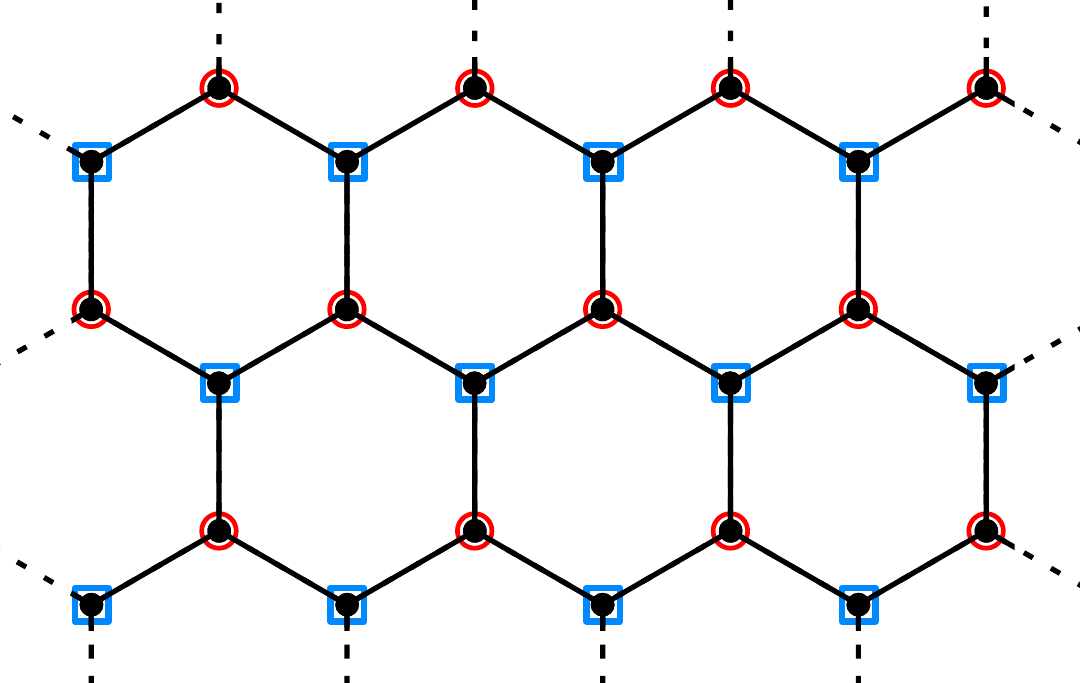}
\caption{In an $n$-site honeycomb lattice there are $n/2$ hexagons, which can be identified by their topmost site (red circles). Half of all lattice sites are not the topmost site of any hexagon (blue squares).}
\label{fig:hexagon1st}
\end{figure}

The amount of $k$-sets is $\mathcal O(n)$, as can be seen as follows. Firstly, it cannot be greater than $n$, since each $k$-set is defined by its first site (``first'' according to any ordering, e.g., top to bottom, left to right, etc); with $n$ sites total, the amount of $k$-sets can be no more than $n$. Secondly, there are cases where not every site is the ``first site'' of a $k$-set. This is better seen with an example on the honeycomb lattice (Fig.\ref{fig:hexagon1st}) where the $k$-sets are its hexagons ($k=6$): taking the topmost site of each hexagon as its ``first'', we see that half the sites are not the topmost site of any hexagon, and indeed there are only $n/2$ hexagons in this case. However, because of the repeating nature inherent of a lattice, this may only occur with a given fraction of all sites ($1/2$ in the example), and the $\mathcal O(n)$ amount of $k$-sets holds in general.

\begin{figure*}
\includegraphics[width=\textwidth]{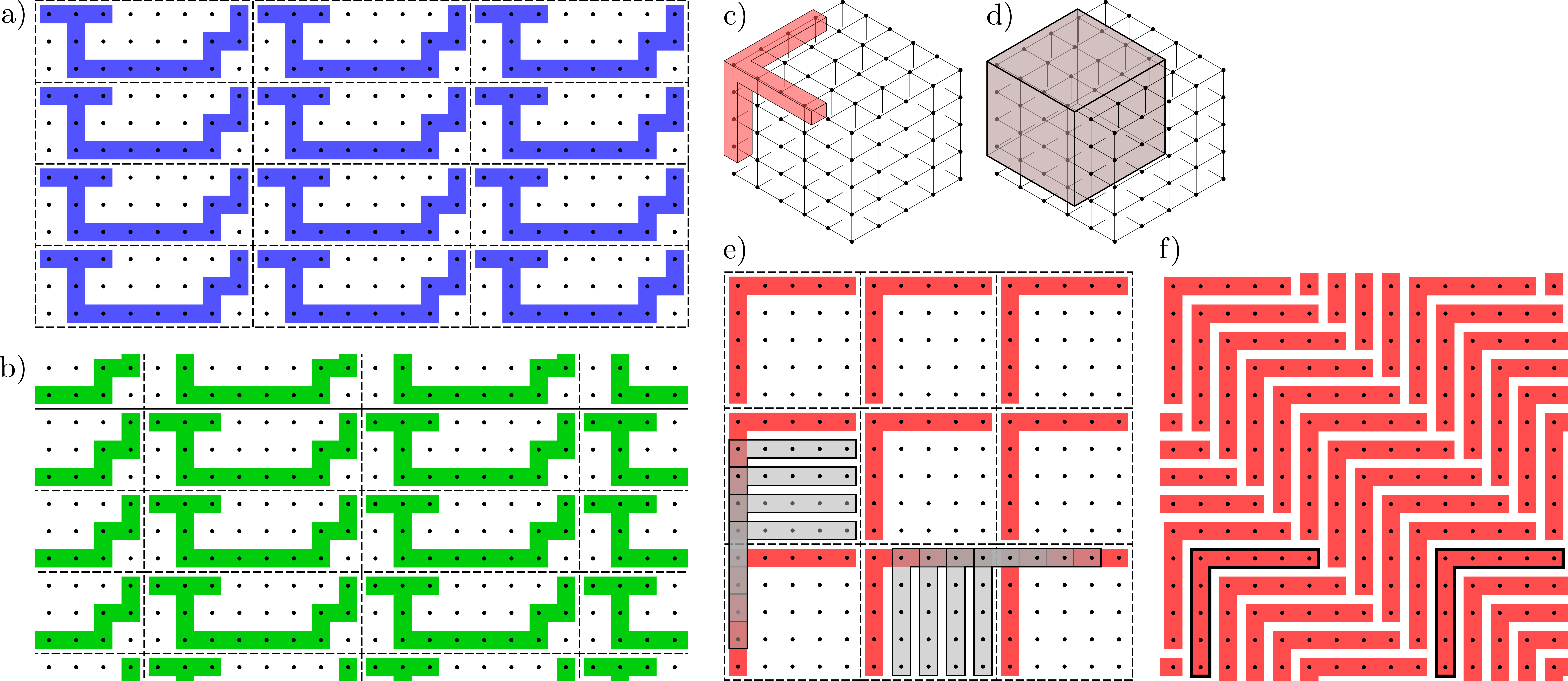}
\caption{a) Tiling of an arbitrary shape by embedding in (non-minimal) lattice cells. The cells (dashed lines) are chosen as the smallest that contain the desired $k$-set (here $k=13$). b) Example of displacement needed to cover $k$-sets overlapping with the ones on the original tiling. Here a displacement of 2 down and 4 to the right is shown; in total $8\times3=24$ displacements are needed [including the original tiling in a)]. At most $k$ displacements in each direction are needed, so at most $k^2$ total for a planar lattice like above, $k^3$ for a three-dimensional lattice. c) Example of a $k$-set shape (here, $k=10$) requiring $\mathcal O(k^3/3)$ displacements in the cubic lattice. d) Illustration of one cell for the scenario of c). 
e) Tiling of an $L$-shaped $k$-set with $k=9$ by embedding in $\frac{k+1}2$-wide square cells. In red, the sets that fit the drawn cells; in gray some of the displaced $L$-shapes, $\frac{k+1}2$ down and $\frac{k+1}2$ to the right, a total of $(k+1)^2/4=25$ displacements (including the original). f) More efficient tiling of the same $L$-shapes. In this case only $k=9$ displacements are necessary in total: a $(k\!+\!1)$-th displacement takes the pattern back to itself, as seen best by the black contoured shape on the left being displaced exactly onto the one on the right after $k+1$ steps.}
\label{fig:tiling}
\end{figure*}

A brute-force approach would be to measure these sets independently. 
Complete information on one $k$-set requires $\mathcal O(3^k)$ measurement settings, to reach all $\mathcal O(n)$ $k$-sets, $N=\mathcal O(n3^k)$ settings are needed. An appropriate tiling, however, reduces $N$. 
Any lattice presents an intrinsic tiling of its sites into minimal cells, but can also be tiled into larger cells each composed of several minimal cells. As such, it is always possible to embed each $k$-set shape in a cell. A suitable form of tiling for our measurements is precisely one where each $k$-set shape fits in one cell, see Fig.~\ref{fig:tiling}a). All such $k$-sets belonging to different cells can be measured simultaneously, consuming $\mathcal O(3^k)$ measurement settings. To cover the $k$-sets that overlap with the ones in the first measurement settings, one must displace the cells in the symmetry directions of the lattice, as in Fig.~\ref{fig:tiling}b). This procedure will require at most $\mathcal O(k^3)$ displacements [see Figs.~\ref{fig:tiling}c) and \ref{fig:tiling}d); for surface lattices, at most $\mathcal O(k^2)$; for linear ones, at most $\mathcal O(k)$], hence a total of $N=\mathcal O(k^33^k)$ measurement settings. 
In conclusion, an appropriate tiling of the $n$-lattice allows to reduce the number of measurement settings to obtain its $k$-sets from $N=\mathcal O(n3^k)$ to at most $N=\mathcal O(3^k)$, independent of $n$. If a fixed number of rotations of the $k$-set is targeted, this simply incurs in constant-prefactor increase in the number of $k$-sets and of necessary cells (e.g.\ if rotations of the shape in Fig.~\ref{fig:tiling}a in 0º, 90º, 180º, and 270º are targeted, this prefactor is $4$). 

Given that the number of repetitions $M$ in our case scales as $M\sim\frac2{\varepsilon^2}[k+\log n+\log(1/\delta)]$ (Section~\ref{sec:numrepetitions}), a comparison with the results in \cite[Theorem 3]{Evans2019} and in \cite[Theorem 1]{Huang2020} is in order. The theorem in \cite{Evans2019} offers an algorithm to estimate the expectation values of $m$ strings of $k$ Pauli operators with $N\sim \frac1{\varepsilon^2}3^k \log m$ independent, randomly selected measurement settings, and without repetitions for statistics (effectively $M=1$). The scheme in \cite{Huang2020}, when employed with randomized measurements on the Pauli bases for these expectation values, similarly yield $N = \mathcal O (\frac1{\varepsilon^2} 3^k\log n)$ with $M=1$. Applying these results to a $k$-RDM on $n$ qubits, one has $m=3^k\binom nk$, hence $NM\sim \mathcal O(\frac1{\varepsilon^2}3^k\log n)$.

Theorem 2 of \cite{Huang2020} applied to this case shows that the product $NM$ is in fact lower-bounded by the same expression, which is also reached in the present work. There is, however, has an important difference: 
we have a lower number of different measurement settings $N\sim\mathcal O(3^k)$ (system-size independent), and a higher number $M\sim\log n$ of repetitions of those for statistics. For experimental considerations, it is clear that repeating the same measurement setting several times is much more amenable than making different ones.

Furthermore, although we can reach this lower bound with the generic embedding presented above, notice that it is not necessarily optimal. Consider the L-shaped $k$-set in Fig.~\ref{fig:tiling}e), with $(k+1)/2$-long arms. The tiling by embedding described above would require $(k+1)/2$ displacements to the left and $(k+1)/2$ down, totaling $(k+1)^2/4=\mathcal O(k^2)$ displacements of the original tiling. A more efficient tiling is shown in Fig.~\ref{fig:tiling}f), and requires only $k$ displacements total.

This tiling procedure demonstrates that the number of measurement settings in local overlapping tomography is independent of the system size. 
Note that in a typical experimental implementation each setting requires $\mathcal O(n)$ detectors, which is still size dependent. Nonetheless, such a dependence of the number of detectors with the system size also holds for the other available procedures in the literature, and an amount of settings which is system-size independent represents a valuable advantage.

The tiling above is, however, typically sub-optimal for a given lattice because of its generality. So we now turn to specific scenarios that are usually encountered in existing setups, and show more efficient measurement strategies for each of them.

\subsection{Qubits in a lattice}
\label{sec:qubits}
 
The simplest case to consider is that of the $2$-RDM for first neighbors. We need that, whenever $i,j$ are neighbors, $\langle\sigma_a^{(i)}\otimes\sigma_b^{(j)}\rangle$ cover all combinations of $(a,b)$. This reduces to the problem of vertex coloring of a graph: neighboring sites must be assigned different labels (colors), and from the number $c$ of colors one obtains the number of measurement settings $N=3^c$ (since each color has to cycle through $x,y,z$). Graph theory shows~\cite{Kubale2004,Beineke2015} that for a graph of degree $\Delta$ (i.e.\ where a site has at most $\Delta$ first neighbors), it is always possible to color it with $\Delta+1$ colors. This gives an upper bound of $N\leqslant3^{\Delta+1}$. Importantly, this graph-theory bound on $c$ can be far from tight~\cite[eq.~(1.2)]{Kubale2004}, hence the same holds for the bound on $N$; better bounds can be written in terms of minimum degrees~\cite[Theorem 2.2]{Beineke2015}. 

Let us now focus on relevant geometries, in which we can obtain bounds that are tighter and hold beyond $k=2$.

\subsubsection{Straight qubit lattices: strings, squares, cubes} 
\label{sec:qubit_straight}

\begin{figure}[tb!]
\includegraphics[width=\columnwidth]{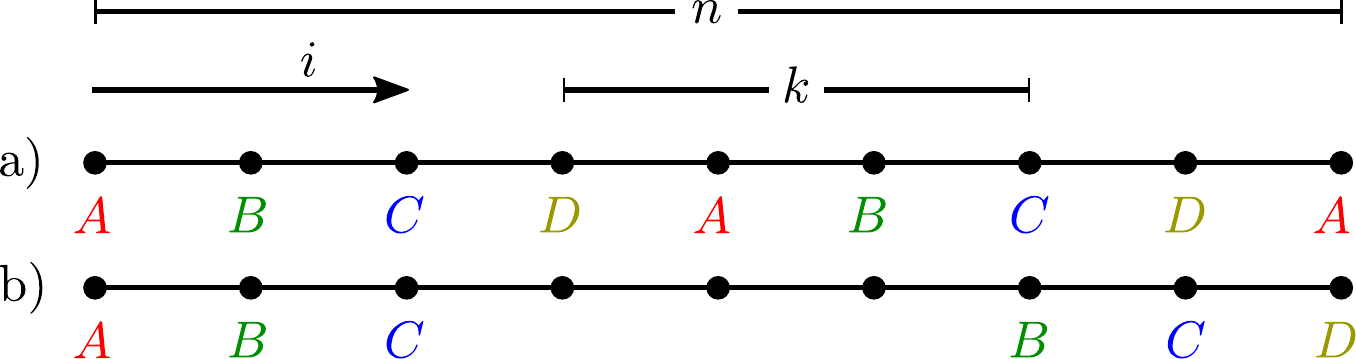}
\caption{Labelings for string lattice [a)] and ring lattice [a) and b)] for $n=9$, $k=4$. For clarity, the numerical labels in the text have been converted to letters with the standard mapping $0\to A$, $1\to B$, $\cdots$, and also color-coded.}
\label{fig:string}
\end{figure}

We begin by the simplest lattice, a string of $n$ qubits, whose results will serve as a building block for the others. The $k$-RDM of neighboring sites on a string are recovered from expectation values of the form
\begin{equation}
\langle\sigma_{a_i}^{(i)}\otimes\sigma_{a_{i+1}}^{(i+1)}\otimes\cdots\otimes\sigma_{a_{i+k-1}}^{(i+k-1)}\rangle \ .
\label{eq:kRDMqubitgeneralform}
\end{equation}
The measurement scheme is defined using a single labeling $\boldsymbol a$ that cycles through a $k$-alphabet, or
\begin{equation}
a_i=i \bmod k
\label{eq:stringlabel}
\end{equation}
[see Fig.~\ref{fig:string}a)]. 
It assigns measurements in the sense that each label runs independently through $x,y,z$, totaling $N=3^k$ measurement settings. 
It is straightforward to see that in this scheme any $k$ consecutive qubits have $k$ different labels, and hence are measured in all combinations of the bases.

It is also clear that, for this or any other $k$-RDM, $k$ is the minimum amount of labels needed: if $k-1$ or less labels were used, any group of $k$ qubits would have coinciding labels, and the qubits assigned the same label would always be measured in the same basis, amounting to an incomplete measurement. Importantly, $N=3^k$ does not scale with the system size $n$.

Another one-dimensional lattice is the ring. A ring here is simply a string whose endpoints are considered first neighbors. A ring of qubits requires extra measurement settings, compared to the open string, when $n$ is not divisible by $k$. A single extra labeling $\boldsymbol b$ is needed (assuming $n\geqslant2k-2$); it labels the first $k-1$ qubits as before, the last $k-1$ qubits with the last $k-1$ letters of the $k$-alphabet, and the remaining qubits are not labeled (i.e.\ not measured). Formally,
\begin{equation}
b_i= \begin{cases} a_i \ , & \ \mathrm{for}\  i\in[k-1]\\ a_{i-n} \ , & \ \mathrm{for}\ i\in\mathrm{last}_{k-1}[n]  \\ \mbox{no label}  & \ \mbox{otherwise} \ , \end{cases}
\label{eq:ringlabel}
\end{equation}
where $\mathrm{last}_{k-1}[n]:=[n]\backslash[n-(k-1)]$ and $a_i$ obeys Eq.~\eqref{eq:stringlabel}. This labeling is exemplified in Fig.~\ref{fig:string}b). Notice that it correctly covers any set of $k$ neighbors across the edge. With two labelings, the number of measurement settings rises to $N=2\times3^k$.

\begin{figure}[tb]
\includegraphics[width=.7\columnwidth]{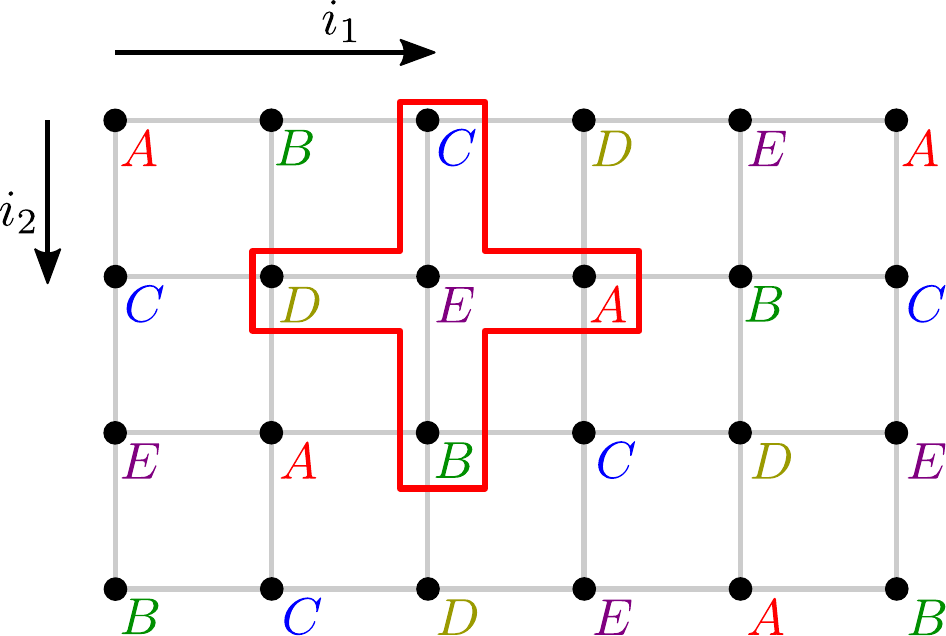}
\caption{Labeling to measure star RDM on the planar square lattice, given by Eq.~\eqref{eq:star2Dlabel}. For clarity, the numerical labels in the text have been converted to letters with the standard mapping $0\to A$, $1\to B$, $\cdots$, and color-coded.}
\label{fig:2Dstar}
\end{figure}

We now move on to two-dimensional geometries, starting by the square lattice. The first RDM tackled here is the star, composed of one qubit and all its neighbors, a $k=5$-RDM in this geometry. A single labeling $\boldsymbol a$ covers all possibilities with the minimum of $k=5$ labels. It labels rows cycling through a 5-alphabet like the string above, but adding an offset of $2$ for each row (see Fig.~\ref{fig:2Dstar}). With lattice sites now denoted by the pair $(i_1,i_2)$, its elements read
\begin{equation}
a_{i_1,i_2}=(i_1 +2i_2)\bmod5 \ .
\label{eq:star2Dlabel}
\end{equation}
Since $5$ labels are used, $N=3^5$ measurement settings are needed.

\begin{figure}[tb]
\includegraphics[width=\columnwidth]{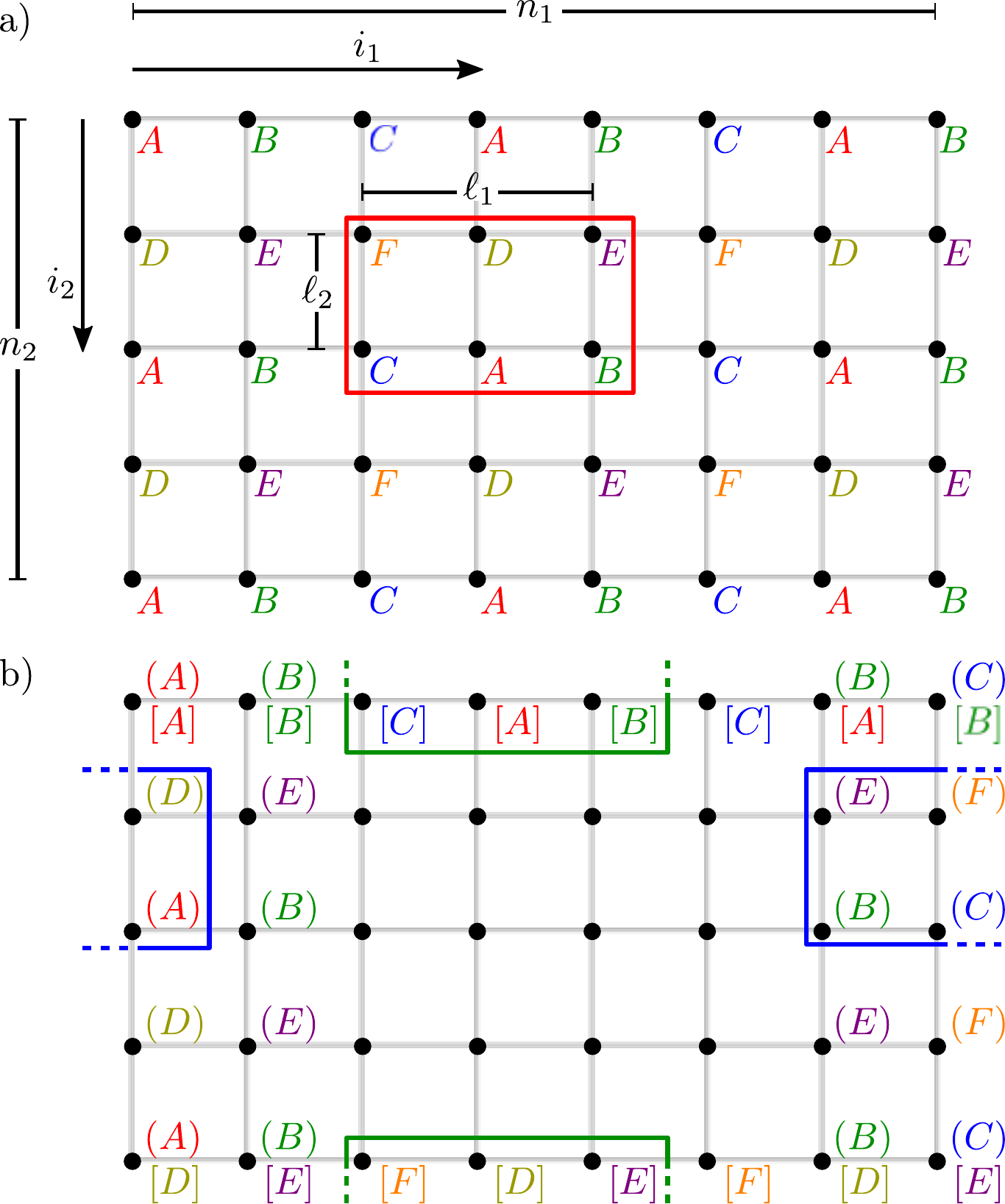}
\caption{a) Labeling $\boldsymbol a$ for $(\ell_1,\ell_2)$-plaquettes in the planar square qubit lattice, from Eq.~\eqref{eq:plaquettelabel}, with $(\ell_1,\ell_2)=(3,2)$. b) Additional labelings needed for the cylinder and torus geometries, with $n_1=8$, $n_2=5$. The cylinder geometry requires $\boldsymbol b$ from Eq.~\eqref{eq:cylinderlabel} (in parentheses); the torus geometry requires $\boldsymbol b$ and also $\boldsymbol c$ from Eq.~\eqref{eq:toruslabel} [in square brackets]. Plaquettes crossing each edge are illustrated. For clarity, the numerical labels in the text have been converted to letters with the standard mapping $0\to A$, $1\to B$, $\cdots$, and color-coded.}
\label{fig:square}
\end{figure}

An $(\ell_1,\ell_2)$-plaquette is defined as an $(\ell_1\times \ell_2)$-qubit rectangle aligned to the lattice (see Fig.~\ref{fig:square}). A measurement scheme for the $k$-RDMs of all $(\ell_1,\ell_2)$-plaquettes (with $k=\ell_1\ell_2$) can be made by composing the previous string scheme. The first row is labeled with an $\ell_1$-alphabet as in the string, the second row is labeled with a different $\ell_1$-alphabet, and the same goes for the first $\ell_2$ rows, each with their different $\ell_1$-alphabets. Row labelings repeat cyclically from the $(\ell_2+1)$-th row onwards, in a total of $\ell_1\ell_2$ labels (Fig.~\ref{fig:square}). Formally, the labeling reads
\begin{equation}
a_{i_1,i_2}=(i_1 \bmod\ell_1)+\ell_1(i_2\bmod\ell_2) \ ,
\label{eq:plaquettelabel}
\end{equation}
and appropriately covers all $(\ell_1,\ell_2)$-plaquettes with the minimal number $k=\ell_1\ell_2$ of labels, hence $N=3^k=3^{\ell_1\ell_2}$.

We call cylinder a square lattice where sites on the last column are first neighbors to same-row sites on the first column. Let the total lattice have dimension $n_1\times n_2$, where $n=n_1n_2$. To measure all $(\ell_1,\ell_2)$-plaquettes in the cylinder when $n_2\bmod\ell_2\neq0$, it takes an additional labeling as in the ring case. The additional labeling is $\boldsymbol b$, with elements
\begin{equation}
b_{i_1,i_2}=
\begin{cases}	a_{i_1,i_2} \ ,& \mathrm{for}\ i_1\in[\ell_1-1] \\ a_{i_1-n_1,i_2} \ ,& \mathrm{for}\ i_1\in\mathrm{last}_{\ell_1-1}[n_1]\\
\mbox{no label} & \mbox{otherwise} \ ,\end{cases}
\label{eq:cylinderlabel}
\end{equation}
with $a_{i_1,i_2}$ from Eq.~\eqref{eq:plaquettelabel}. It labels only the leftmost and rightmost $(\ell_1-1)$ columns and, analogously to the ring, covers all $(\ell_1,\ell_2)$-plaquettes across the vertical edges. With two labelings, the number of measurement settings is $N=2\times3^k$.

When the first and last rows of a cylinder are also taken as first neighbors (column-wise), we obtain a torus. If both $n_1\bmod\ell_1\neq0$ and $n_2\bmod\ell_2\neq0$, a third labeling $\boldsymbol c$ is needed, analogous to Eq.~\eqref{eq:cylinderlabel}, but switching horizontal and vertical directions. With $a_{i_1,i_2}$ from Eq.~\eqref{eq:plaquettelabel}, $\boldsymbol c$ is given by
\begin{equation}
c_{i_1,i_2}=
\begin{cases}	a_{i_1,i_2}\ ,& \mathrm{for}\ i_2\in[\ell_2-1] \\ a_{i_1,i_2-n_2} \ ,& \mathrm{for}\ i_2\in\mathrm{last}_{\ell_2-1}[n_2]\\
\mbox{no label} & \mbox{otherwise} \ ,\end{cases}
\label{eq:toruslabel}
\end{equation}
and labels the topmost and bottommost $(\ell_2-1)$ rows. With three labelings, the number of measurement settings is $N=3\times3^k=3^{k+1}$. Notably, the torus topology makes a third labeling necessary in general (Appendix~\ref{sec:creases}). 

In fact, any labeling $\boldsymbol a$ of the square lattice can be patched for the cylinder and torus geometries with the strategy above, as long as the maximal dimensions of the desired RDM are $\ell_1\times\ell_2$. Applying this to the star-RDM labeling --- i.e.\ substituting Eq.~\eqref{eq:star2Dlabel} in Eqs.~\eqref{eq:cylinderlabel},\eqref{eq:toruslabel} with $\ell_1=\ell_2=3$ ---, one finds
\begin{align}
b_{i_1,i_2}=&\begin{cases}	(i_1+2i_2)\bmod5 \ ,& \mathrm{for}\ i_1\in[2] \\ (i_1-n_1+2i_2)\bmod5 \ ,& \mathrm{for}\ i_1\in\mathrm{last}_{2}[n_1]\\
\mbox{no label} & \mbox{otherwise}		\end{cases} \label{eq:starcylinder}\\
c_{i_1,i_2}=&\begin{cases}	(i_1+2i_2)\bmod5 \ ,& \mathrm{for}\ i_2\in[2] \\ (i_1+2i_2-2n_2)\bmod5 \ ,& \mathrm{for}\ i_2\in\mathrm{last}_{2}[n_2]\\
\mbox{no label} & \mbox{otherwise}\end{cases}
\label{eq:startorus}
\end{align}
and the number of measurement settings is $N=2\times3^5$ for the cylinder ($\boldsymbol a$ and $\boldsymbol b$) and $N=3\times3^5=3^6$ for the torus geometry ($\boldsymbol a$, $\boldsymbol b$ and $\boldsymbol c$).

We now move on to a three-dimensional, cubic lattice. We begin by the star RDM, which includes a qubit and all its (six) first neighbors.  This $k=7$-RDM can be realized with one labeling $\boldsymbol a$ composed of $k=7$ labels; it labels each plane as in the 2D star configuration (but with a 7-alphabet), with an offset of $3$ between successive planes. With each site labeled by $(i_1,i_2,i_3)$, $\boldsymbol a$ reads
\begin{equation}
a_{i_1,i_2,i_3}=(i_1+2i_2+3i_3)\bmod7
\label{eq:star3Dlabel}
\end{equation}
and implies $N=3^7$ measurement settings.

\begin{table}[tb]
\begin{tabular}{|c|c|c|c|}\hline
Geometry			& Labelings & \shortstack{$N$ (No. of \\measmts)} & Eqs.\\ \hline
String				& 1 &	$3^k$&\eqref{eq:stringlabel}\\
Ring	(1D)		& 2 &	$2\times3^k$&\eqref{eq:stringlabel},\eqref{eq:ringlabel}\\\hline
Star (plane)	&	1	&	$3^5$&\eqref{eq:star2Dlabel}\\
$(\ell_1,\ell_2)$-plaquette&	1 & $3^k=3^{\ell_1\ell_2}$&\eqref{eq:plaquettelabel}\\
Cylinder			&	$2\times$(planar)				&	$2\times$(planar)&\eqref{eq:plaquettelabel},\eqref{eq:cylinderlabel}\\
Torus					&	$3\times$(planar)				&	$3\times$(planar)&\eqref{eq:plaquettelabel},\eqref{eq:cylinderlabel},\eqref{eq:toruslabel}\\\hline 
Star (cubic)			&	1	&	 $3^7$&\eqref{eq:star3Dlabel}\\
$(\ell_1,\ell_2,\ell_3)$-block&	1	&	$3^k=3^{\ell_1\ell_2\ell_3}$&\eqref{eq:blockettelabel}\\
Thick ring (3D)	&	$2\times$(cubic)	&	$2\times$(cubic)&\eqref{eq:blockettelabel},\eqref{eq:thickringlabel}\\\hline
\end{tabular}
\caption{Summary of the number of labelings and measurement settings for $k$-RDMs on the geometries considered in Subsection~\ref{sec:qubit_straight}. For comparison, previous methods in the literature require at least $N\sim e^{\mathcal O(k)}\log n$ different measurement settings. The values of the cyclic geometries (ring, cylinder, torus, thick ring) are upper bounds, saturated when the dimensions of the lattice are not divisible by those of the RDM.}
\label{tab:summarySquare}
\end{table}

An $(\ell_1,\ell_2,\ell_3)$-block is an $(\ell_1,\ell_2,\ell_3)$-qubit parallelepiped aligned to the lattice. Analogously to the plaquette, all such blocks can be labeled by
\begin{equation}
a_{i_1,i_2,i_3}=(i_1 \bmod\ell_1)+\ell_1(i_2\bmod\ell_2)+\ell_1\ell_2(i_3\bmod\ell_3) \ ,
\label{eq:blockettelabel}
\end{equation}
with $N=3^k=3^{\ell_1\ell_2\ell_3}$ measurement settings.

If the rightmost face of the cubic lattice is considered first neighbor of the leftmost face, we then have a ``thick'' ring. With the lattice having dimensions $n_1\times n_2\times n_3$, and taking the $i_1=0$ face as neighbor to the $i_1=(n_1-1)$ face, this thick ring has a cross section of dimension $n_2\times n_3$ qubits. As in the original one-dimensional ring, this geometry requires one additional labeling, given by
\begin{equation}
b_{i_1,i_2}=
\begin{cases}	a_{i_1,i_2,i_3} \ ,& \mathrm{for}\ i_1\in[\ell_1-1] \\ a_{i_1-n_1,i_2,i_3} \ ,& \mathrm{for}\ i_1\in\mathrm{last}_{\ell_1-1}[n_1]\\
\mbox{no label} & \mbox{otherwise} \ ,\end{cases}
\label{eq:thickringlabel}
\end{equation}
totaling $2\times3^{\ell_1\ell_2\ell_3}$ measurement settings. As was done for the cylinder and torus in the two-dimensional lattice, one can consider the other faces of the three-dimensional lattice to be first neighbors, and the additional labelings follow analogously.

We summarize the results of this Subsection in Table~\ref{tab:summarySquare}.

\subsubsection{Further qubit lattices: triangle and honeycomb}
\label{sec:qubit_further}

\begin{figure}
\includegraphics[width=.8\columnwidth]{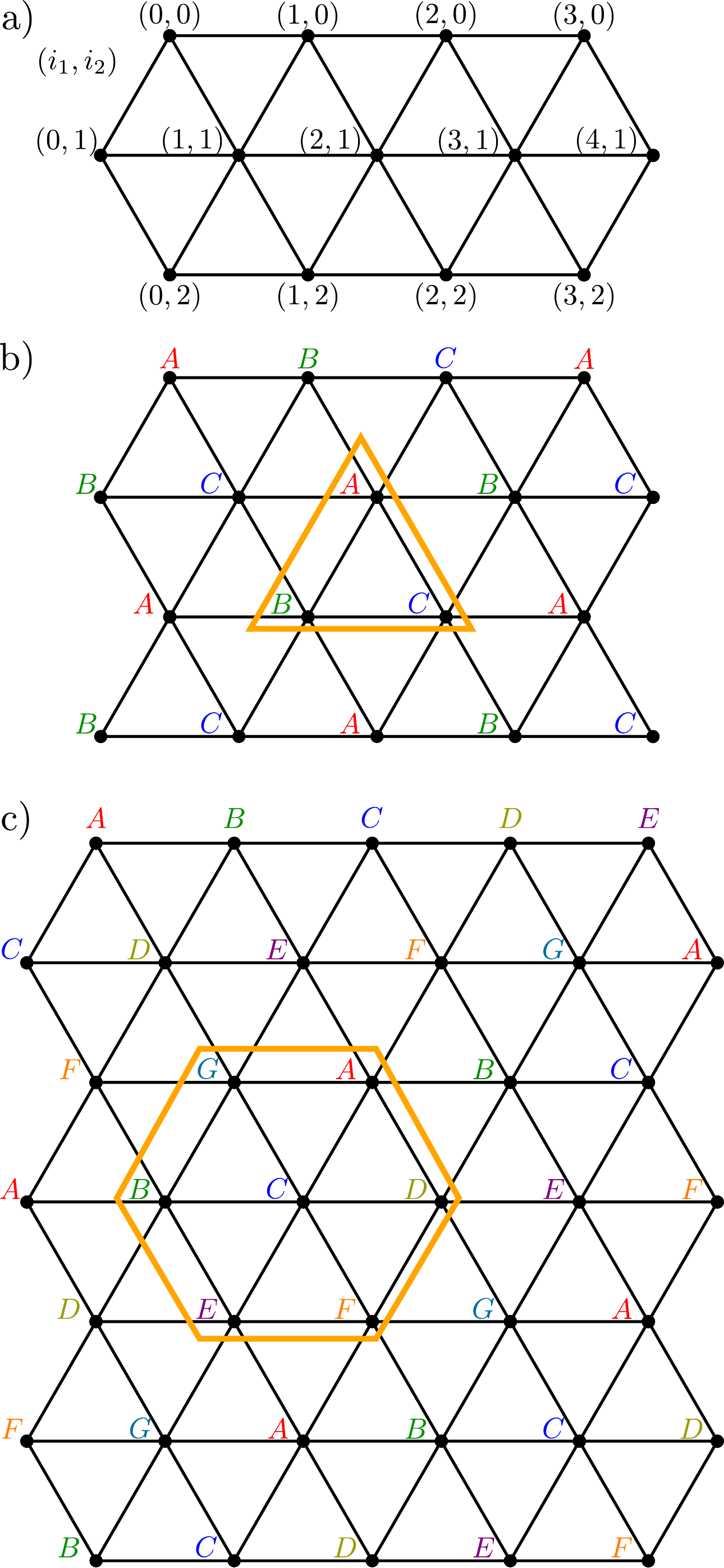}
\caption{Labeling of the triangle lattice. a) Numbering scheme used in Eqs.~\eqref{eq:triangleRDM}, \eqref{eq:trianglestarRDM}. b) Three-label scheme for the (triangular) plaquette. c) Seven-label scheme for the star RDM (qubit and its neighbors).}
\label{fig:triangle}
\end{figure}

In this Subsection we treat different geometries, namely the triangle and the honeycomb lattices.

For the triangle geometry, we begin with a labeling scheme that covers triangle RDMs. With three labels, it is also the minimal scheme for 2-RDMs. With the numbering defined in Fig.~\ref{fig:triangle}a), this labeling reads
\begin{equation}
a_{i_1,i_2} = (i_2\bmod2+i_1)\bmod3
\label{eq:triangleRDM}
\end{equation} 
and is illustrated in Fig.~\ref{fig:triangle}b). 

Additionally, we present a scheme for the star RDM on this geometry, composed of a qubit and all its six neighbors (hence a $7$-RDM):
\begin{equation}
a_{i_1,i_2} = \begin{cases} (i_1 - i_2) \bmod7 \ ,& \mbox{for $i_2$ even}\\
														(i_1 - i_2 + 3) \bmod7 \ ,& \mbox{for $i_2$ odd}	\end{cases}
\label{eq:trianglestarRDM}
\end{equation}
which is illustrated in Fig.~\ref{fig:triangle}c).

\begin{figure*}
\includegraphics[width=2\columnwidth]{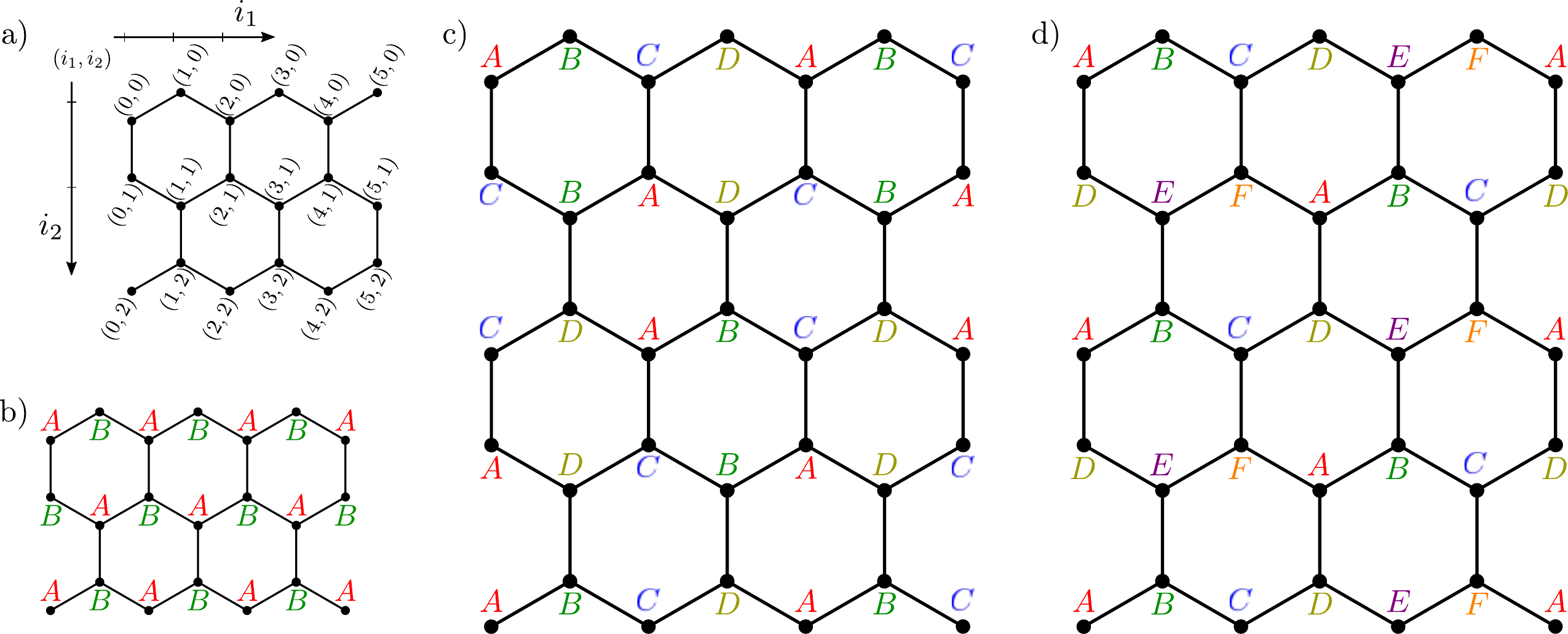}
\caption{Labeling of the honeycomb lattice for different RDMs.
 a) Explicit illustration of the numbering used in Eqs.~\eqref{eq:honeycomb2RDM}, \eqref{eq:honeycomb4labels}, \eqref{eq:honeycomb6labels}. b)~Two-label scheme for first-neighbor $2$-RDMs, Eq.~\eqref{eq:honeycomb2RDM}. 
c)~Four-label scheme, suitable for $3$-RDMs and any $4$-RDM where the four qubits are contiguous, but not in the same hexagon, given by Eq.~\eqref{eq:honeycomb4labels}.  d)~Six-label scheme, suitable for $4$-, $5$-, $6$-RDMs where all qubits of the RDM belong to same hexagon, given by Eq.~\eqref{eq:honeycomb6labels}. 
 For clarity, the numerical labels in the text have been converted to letters with the standard mapping $0\to A$, $1\to B$, $\cdots$, and color-coded.}
\label{fig:honeycomb}
\end{figure*}

We now deal with qubits on a honeycomb lattice. A first-neighbor $2$-RDM can be achieved with two labels, simply assigning neighbors different labels. With the numbering exemplified in Fig.~\ref{fig:honeycomb}a), this labeling reads 
\begin{equation}
a_{i_1,i_2}=i_1+i_2\bmod2 \ ,
\label{eq:honeycomb2RDM}
\end{equation}
and is shown in Fig.~\ref{fig:honeycomb}b).

Now we show a labeling that is useful for different sets of neighbors. It contains 4 different labels, as seen in Fig.~\ref{fig:honeycomb}c), and covers any $3$-RDM, as well as any $4$-RDM where the four qubits do not all belong to same hexagon. This includes the star configuration (a qubit and its three neighbors), as well as 4 qubits along the rows, among others:
\begin{equation}
a_{i_1,i_2} = \begin{cases} (-1)^{i_2}\ i_1    &\mbox{for $i_2\bmod4 = 0,3$}\ ,\\
														(-1)^{i_2}\ i_1 +2 &\mbox{for $i_2\bmod4 = 1,2$}\ , \end{cases}
\label{eq:honeycomb4labels}
\end{equation}
equivalently, $a_{i_1,i_2} = \left((-1)^{i_2}i_1\!+2\!\left\lfloor\!\frac{(i_2+1)\bmod4}{2}\!\right\rfloor\right)\bmod4$.
It is perhaps clearer to describe this labeling with the following algorithm: for the first row, assign labels sequentially ($a_{i_1,0}=i_1\bmod4$). For the remaining sites, assign them the same color as the site diametrically opposite to it in any given hexagon, as in Fig.~\ref{fig:honeycomb}c).

This is evidently the minimal labeling for the $4$-RDMs, and is also the minimal for the $3$-RDMs, which cannot be covered with 3 labels only.

Lastly, we present a final labeling for the honeycomb, composed of 6 labels. It covers hexagon RDMs (containing all qubits in a hexagon), as well as 4-RDMs and 5-RDMs with all qubits in the same hexagon. It is depicted in Fig.\ref{fig:honeycomb}d) and is given by
\begin{equation}
a_{i_1,i_2} = [3 (i_2\bmod2)+i_1]\bmod6 \ .
\label{eq:honeycomb6labels}
\end{equation}
Interestingly, this labeling also covers (non-optimally) all previously mentioned honeycomb RDMs.

\begin{table}
\begin{tabular}{|c|c|c|c|}\hline
No. of qubits & Labels & \shortstack{$N$ (No. of \\measmts)}&Eq.\\\hline
\multicolumn{4}{|c|}{Triangle}\\\hline
$3$ (plaquette)&$3$&$3^3$&\eqref{eq:triangleRDM}\\
$7$ (star RDM) &$7$&$3^7$&\eqref{eq:trianglestarRDM}\\\hline
\multicolumn{4}{|c|}{Honeycomb}\\\hline
$2$&2&$3^2$&\eqref{eq:honeycomb2RDM}\\
$3$, or $4$ not in same hexagon&4&$3^4$&\eqref{eq:honeycomb4labels}\\
up to $6$ in single hexagon&$6$&$3^6$&\eqref{eq:honeycomb6labels}\\\hline
\end{tabular}
\caption{Summary of the number of labels and measurement settings for neighbor $k$-RDMs on the sets and geometries considered in Subsection~\ref{sec:qubit_further}. For comparison, previous methods in the literature require at least $N\sim e^{\mathcal O(k)}\log n$ different measurement settings.}
\label{tab:summaryfurther}
\end{table}

We summarize the results of this Section in Table~\ref{tab:summaryfurther}.

\subsection{Fermions in a lattice}
\label{sec:fermions}

For fermions in a lattice, $k$-RDMs provide information on arbitrarily distant lattice sites, which is not the purpose of this work. We are interested in $\langle\gamma_{i_0}\gamma_{i_1}\cdots\gamma_{i_{2k-1}}\rangle$, where all $\gamma_i$ belong to a subset of lattice sites. We call these expectation values, then, elements of a \emph{lattice-restricted matrix} of size $k$, or $k$-LRM, for short. An entire $k$-LRM will contain such expectation values for all lattice subsets of a certain class, e.g.\ all neighboring pairs of lattice sites.  A $k$-LRM contains less information than a $k$-RDM, unless the subset is taken to be the entire lattice, for which the $k$-LRM coincides with the $k$-RDM. 

Additionally, for fermions one cannot simply rely on the tensor-product structure of operators acting on different sites to ensure commutativity (hence joint measurability). As such, measuring a given element no longer assumes access to individual measurements on each site. Different Majoranas always anticommute, and they can be compounded to form commuting operators --- e.g.\ two pairs of Majoranas commute as long as all four Majoranas involved are different. We call $j$-Majorana string a product $\gamma_p\cdots\gamma_q$ of $j$ Majoranas, and the overlap of two Majorana strings is the set of Majoranas the two have in common. In general, two Majorana strings commute when the length of their overlap has the same parity as the product of their lengths (see Appendix~\ref{sec:overlapcommute}).

The 1-LRMs for neighbor fermions on a lattice present a simplified structure: since two different pairs of Majorana operators can only overlap in a single $\gamma_i$, any overlapping pairs anticommute and are incompatible. The 1-LRM then reduces to the problem of covering all relevant pairs of Majoranas without using any overlap. Interestingly, the problem reduces to that of edge-coloring a graph \cite{Vizing1964,Stiebitz2012}: how to color the edges of a graph such that any two edges that share a vertex have different colors.

\begin{figure}
\includegraphics[width=\columnwidth]{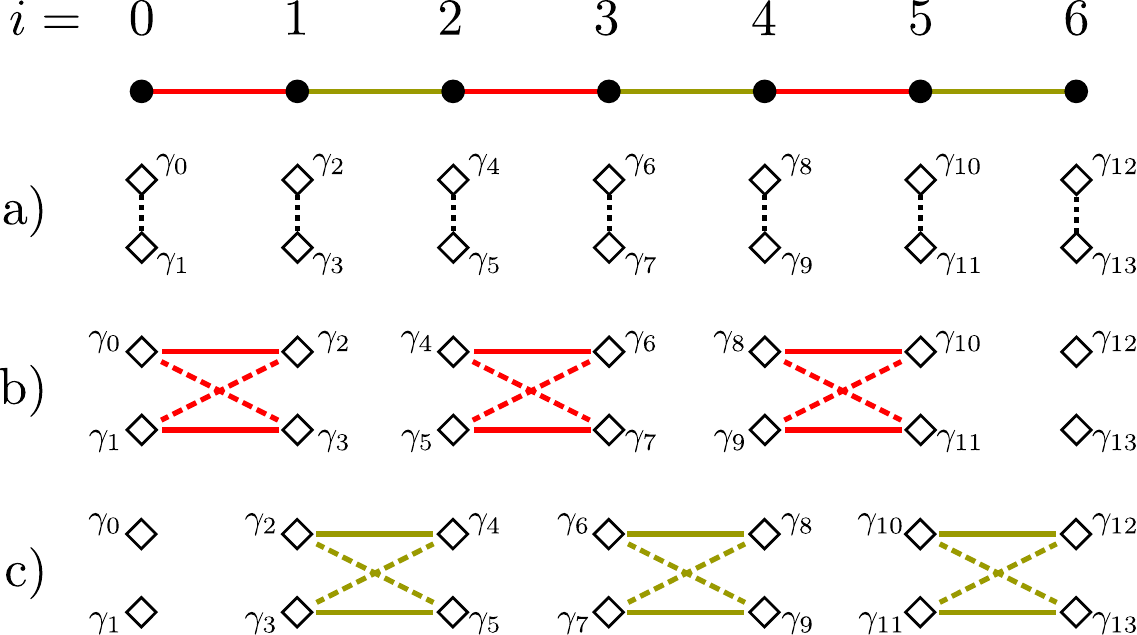}
\caption{Fermionic 1-LRMs, edge-coloring of a graph and Vizing's theorem. Each lattice site (black circle) has $\Delta=2$ neighbors and the graph can be colored with only $c=2$ colors. The scheme shows the $N=2c+1=5$ measurement settings needed on the Majorana operators (white losanges). a) Same-site measurement; b) two measurement settings for the red (dark) edges, the first for equal-parity $\gamma_j$ (solid lines) and the second for opposite-parity $\gamma_j$ (dashed lines); c) two measurement settings for the tan (light) edges, analogous to b).}
\label{fig:1LRM}
\end{figure}

To see this, consider that a graph $G$ defines the lattice, with each fermion site being a vertex and an edge connecting first neighbors. To each site $i$ correspond two Majoranas $\gamma_{2i}$, $\gamma_{2i+1}$. Since the pairs $\gamma_{2i}\gamma_{2i+1}$ for different values of $i$ all commute, they can all be measured at once in a first measurement setting [same-site measurement, see Fig.~\ref{fig:1LRM}a)]. If an edge connects vertices $i$ and $j$, the pairs composed of the elements in $\{\gamma_{2i},\gamma_{2i+1},\gamma_{2j},\gamma_{2j+1}\}$ must be measured. This requires two measurement settings (besides the same-site one): first the equal-parity measurement $\{\gamma_{2i}\gamma_{2j},\gamma_{2i+1}\gamma_{2j+1}\}$ [Fig.~\ref{fig:1LRM}b)], and second the opposite-parity measurement $\{\gamma_{2i}\gamma_{2j+1},\gamma_{2i+1}\gamma_{2j}\}$ [Fig.~\ref{fig:1LRM}c)]. To measure the entire LRM one could, in principle, use two measurement settings for each edge of $G$, but the total amount of settings would be unnecessarily high. A more efficient approach is to make as many of those measurements in parallel as possible. The restrictions to parallelization are that edges connected to the same vertex cannot be measured at the same step: if $(i,j)$ and $(i,j')$ are edges of $G$, their measurements cannot be made in parallel for they would overlap in $\gamma_{2i}$ or $\gamma_{2i+1}$. This restriction is precisely described by the edge-coloring of $G$. To each color correspond two measurement settings, on top of the same-site one. For a total of $c$ colors, this scheme takes $N=2c+1$ measurement settings (see Fig.~\ref{fig:1LRM}).

Importantly, Vizing's theorem \cite{Vizing1964,Stiebitz2012} bounds the amount of colors needed to edge-color a given graph: for a graph of degree $\Delta$, the amount of colors needed is either $c=\Delta$ or $c=\Delta+1$. As such, this scheme makes use of, at most, $N=2\Delta+3$ measurement settings.

Finally, this scheme can be extended to $1$-LRMs that include next-to-nearest neighbors. One need only create an auxiliary graph $G'$ that has all vertices and edges of the original graph $G$, plus edges that connect next-to-nearest neighbors. Edge-coloring of $G'$ indicates a suitable measurement scheme that requires at most $N=2\Delta'+3$ measurement settings, where $\Delta'$ is the degree of $G'$.

The 2-LRM for pairs of first neighbors presents, surprisingly, an even simpler structure.  For any neighboring pair of lattice sites $(i,j)$, the first-neighbor 2-LRM has a single element $\langle\gamma_{2i}\gamma_{2i+1}\gamma_{2j}\gamma_{2j+1}\rangle$. For any other pair $(i',j')$ of lattice sites, the overlap is either of two or zero Majoranas, so all such elements commute. Hence all pertinent Majorana strings are jointly measurable, i.e. a single measurement setting $N=1$ can be made to provide the 2-LRM for pairs of first neighbors.

\begin{figure*}
\includegraphics[width=\textwidth]{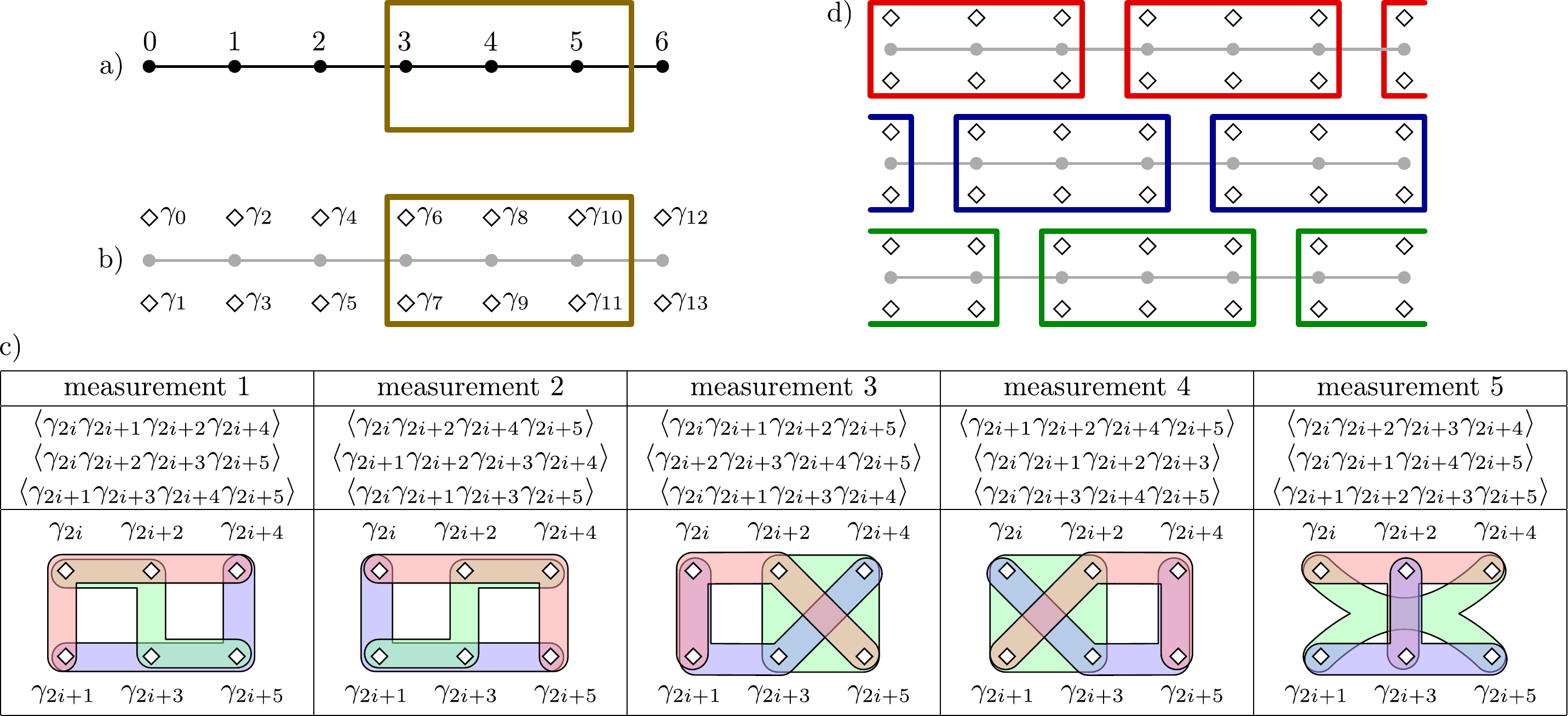}
\caption{Measurement settings of a three-lattice-site tile of a Fermionic 2-LRM. Illustration of the string lattice [black circles, in a)] and of the Majorana operators, two per lattice site [white losanges, in b)]. In c), the five measurement settings for each tile are shown, each obtains three (compatible) elements of the 2-LRM. In d), tiling for the string lattice with three displacements total.}
\label{fig:2LRM3tile}
\end{figure*}

Let us now go beyond first neighbors in specific geometries, beginning with the string lattice.

For the 2-LRM for three contiguous sites in the string lattice, $i$, $i+1$, $i+2$, we will use a tiling strategy. Each tile will be composed of three sites [six Majoranas; see Fig.~\ref{fig:2LRM3tile}a) and b)]. Five measurement settings, with three LRM elements each, suffice to cover a tile; these are shown in Fig.~\ref{fig:2LRM3tile}c). A total of three displacements of the original tiling is needed [see Fig.~\ref{fig:2LRM3tile}d)], which leads to $N=15$ measurement settings total.

\begin{figure}
\includegraphics[width=.65\columnwidth]{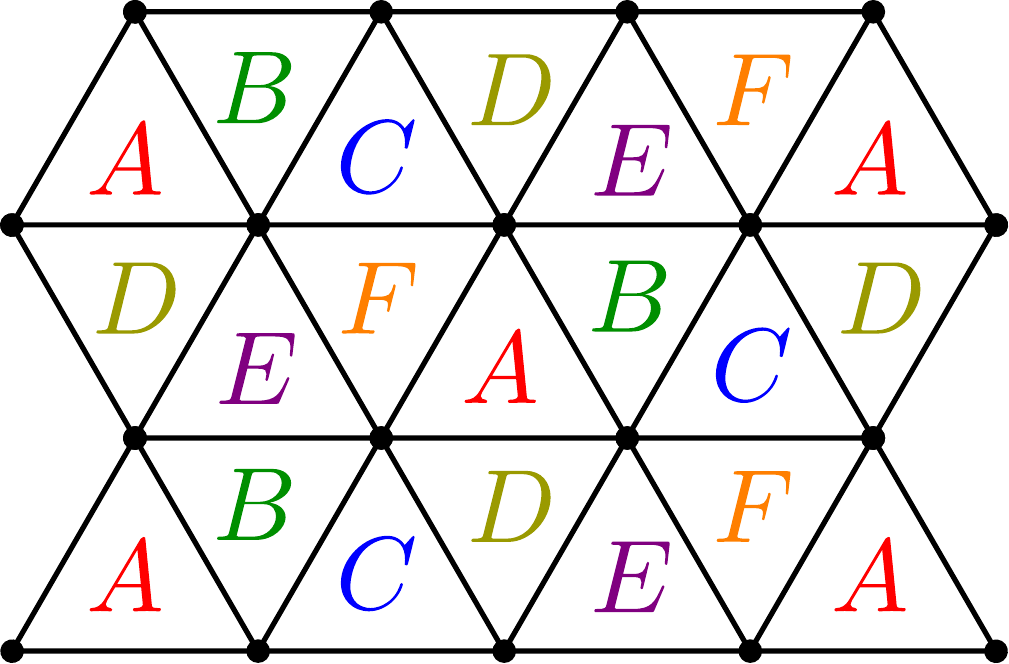}
\caption{Illustration of the tiling for triangle 2-RLM. For a given letter, e.g.\ $A$, each tile comprises the three lattice sites of the triangles assigned that letter. In each tile, measurement settings as in Fig.~\ref{fig:2LRM3tile}c), with all same-letter tiles measured simultaneously. Six tilings, one for each letter, are needed.}
\label{fig:2LRMtriangle}
\end{figure}

In fact, these measurements remain useful for three-site 2-LRMs of any geometry. Consider, e.g., the triangular lattice from Fig.~\ref{fig:2LRMtriangle} and in it the fermionic 2-LRM for sites arranged in a triangle. The lattice needs to be tiled into triangles, and for each triangle, the five measurement settings in Fig.~\ref{fig:2LRM3tile} are used (with a simple change of geometry). The tiling must be such that tiles do not overlap, and a total of six displacements are needed, as shown if Fig.~\ref{fig:2LRMtriangle}. This leads to $N=30$ measurement settings total for this 2-RLM.

\section{Concluding remarks}
\label{sec:conclusion}

The full determination of the quantum state of a many-particle system is practically out of reach when increasing the system size. However, it is possible to access to partial, physically relevant information about this system with a much less demanding effort.  We have studied local overlapping tomography of qubits and fermions in a lattice. We have specifically focused on $k$-RDMs (and $k$-LRMs) that are localized in neighboring sites of the lattice, which are often the subset that draws more interest in theoretical and practical terms. Resorting only to projective, product measurements, we have shown that optimal measurement complexity can be directly tied to graph-coloring for the lowest values of $k$; and we have also studied optimal measurements tailored to several relevant geometries. An important result is that, unlike for the case of full RDMs or even more efficient randomized measurement schemes, the amount of measurement settings to obtain these local matrices no longer depends on system size $n$. This has been seen in the studied geometries, in connection to graph-coloring, and also in the form of a general tiling argument.

Local QOT is a promising candidate for measurements of many-body systems whose correlations are restricted to few bodies, but also as a probe to characterize $k$-local generators of correlations. One should notice that all relevant quantum-computing platforms operate based on 1- and 2-qubit gates acting on neighboring particles, and hence are a potential application of local QOT methods. 
 This opens the path towards turning this form of local overlapping tomography into a viable candidate for measurements in practical scenarios.

\begin{acknowledgments}
We thank X. Bonet-Monroig, A. Zhao, G. Garc\'ia-P\'erez, S. Maniscalco and S. Flammia for helpful discussions and comments. We acknowledge support from the Government of Spain (FIS2020-TRANQI, EU NextGen Funds, Misiones CUCO and Severo Ochoa CEX2019-000910-S), Fundació Cellex, Fundació Mir-Puig, Generalitat de Catalunya (CERCA and QuantumCAT), the ERC AdG CERQUTE, the European Union Horizon 2020 research and innovation program under the Marie Sklodowska-Curie grant agreement No. 754558 (PREBIST), and the AXA Chair in Quantum Information Science.

\end{acknowledgments}

\bibliography{Few-Measurement,ExtraShortNames}

\appendix

\section{Number of repetitions}
\label{sec:ChernoffHoeffding4us}

Here we prove some relations on the number of repetitions $M$ needed for statistical purposes based on the Chernoff-Hoefding bound, as in \cite{Cotler2020}. Consider that $M$ measurements of each variable are made, and that each measurement has $\pm1$ as possible outcomes. Then the Chernoff-Hoefding bound states that the probability of error obeys
\begin{equation}
P(\mathrm{error}_i>\varepsilon) \leqslant 2 e^{-M\varepsilon^2/2} \ ,
\label{eq:ChernoffHoefding}
\end{equation} 
where $\mathrm{error}_i$ is defined as the difference in modulus between the expectation value of the $i$-th variable and its estimate (arithmetic mean of $M$ measured values). In words, it states that the probability of error greater than $\varepsilon$ is exponentially suppressed. The union bound can then be used to bound the probability of high error in any of the variables:
\begin{equation}
P(\mathrm{error}_i>\varepsilon \ \mbox{for any}\ i) \leqslant \sum_i 2 e^{-M\varepsilon^2/2} \ ,
\label{eq:unionbound}
\end{equation}
where the sum in $i$ runs over all measured variables. The full $k$-RDM considered in \cite{Cotler2020} requires measuring $4^k-1$ variables of each of the $\binom nk$ sets of $k$ qubits, or $(4^k-1)\binom nk$ variables total. So for the full $k$-RDM,
\begin{equation}
P(\mathrm{error}_i>\varepsilon \ \mbox{for any}\ i) \leqslant 2 (4^k-1)\binom nk e^{-M\varepsilon^2/2} \ ,
\label{eq:CHfullRDM}
\end{equation}
and for that probability to be smaller than a small value $\delta$ we can set
\begin{equation}
M \sim \frac2{\varepsilon^2} \left(k\log n+\log\frac1\delta\right) \ \ \ \mbox{(full $k$-RDM).}
\label{eq:Mfull}
\end{equation}
In our case, we do not measure all $\binom nk$ sets of $k$ qubits, only $\mathcal O(n)$ of them. We can then set
\begin{equation}
M \sim \frac2{\varepsilon^2} \left(k+\log n +\log\frac1\delta\right) \ \ \ \mbox{(neighbors only).}
\label{eq:Mneighbors}
\end{equation}
For fermions, there are instead $\binom{2n}{2k}$ variables total in the full $k$-RDM, and still $\mathcal O(n)$ for our local $k$-RDMs, and the last two equations still stand. 

\begin{figure}[b]
\includegraphics[width=\columnwidth]{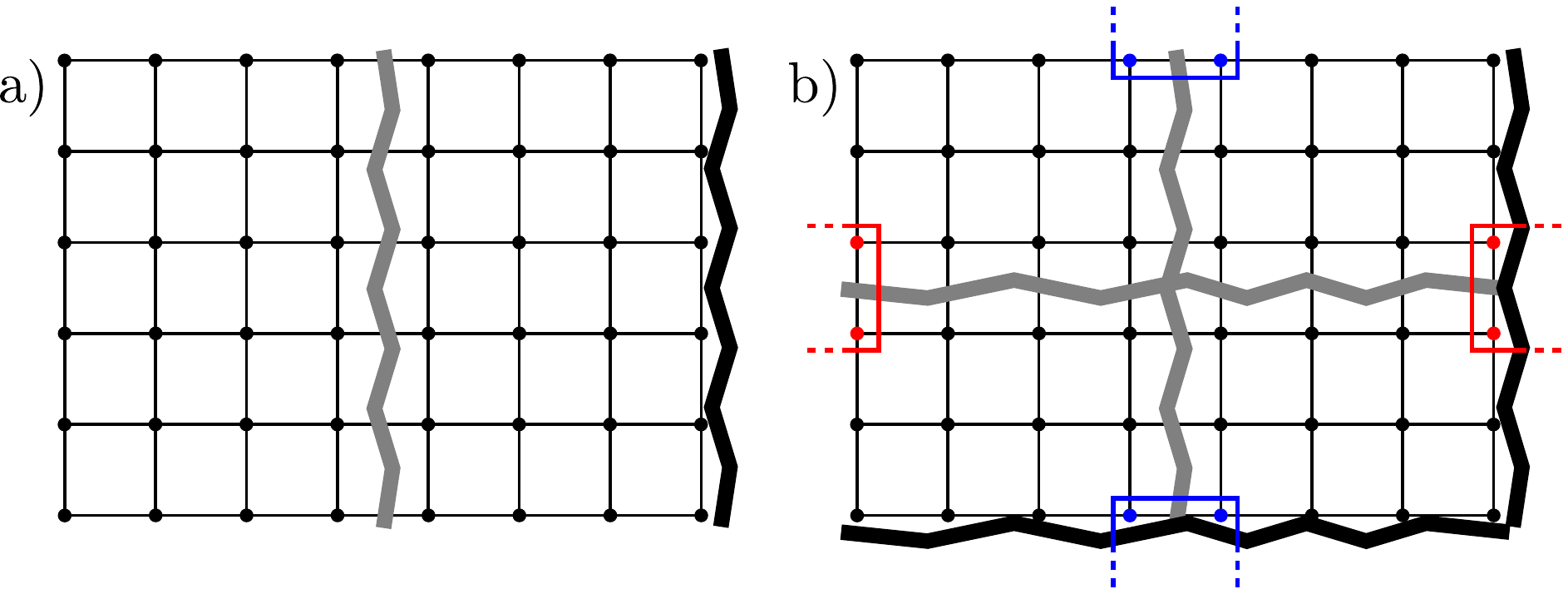}
\caption{Creases on an $8\times6$ square lattice generated by different labelings. Black jagged lines: creases by labeling $\boldsymbol a$ from Eq.~\eqref{eq:plaquettelabel} on a) a cylinder and b) a torus. Gray jagged lines: creases by additional labeling $\boldsymbol b$ on each geometry [in the case of $\ell_1=4$, $\ell_2=3$, from Eq.~\eqref{eq:cylinderlabel} exactly]. Notice how on the cylinder a) the second labeling fully patches the creases of the first, but on the torus b) there are two crossing points between black and gray creases that remain unpatched (the two plaquettes shown cannot be measured with two labelings); the scenario calls for a third labeling.}
\label{fig:creasepatch}
\end{figure}
\section{Patching creases on looping geometries}
\label{sec:creases}
Let us prove the statement that to measure plaquettes on the torus topology a third labeling is in general necessary. This happens because whenever $n_1\bmod\ell_1\neq0$ and $n_2\bmod\ell_2\neq0$ any labeling produces a vertical and a horizontal ``crease'' between lattice sites, across which plaquettes cannot be measured, and the additional labelings are meant to patch these creases (Fig.\ref{fig:creasepatch}). With only two labelings, we have two pairs of creases, each pair composed of a horizontal and a vertical crease. The torus topology ensures that these pairs cross one another in at least two points, though. Any plaquette that encompasses one of the crossing points cannot be covered by the two labelings, requiring a third one.

\section{Majorana overlaps and commutation}
\label{sec:overlapcommute}

Consider that two Majorana strings $s_A$, $s_B$ are composed of substrings $s_j$, $s_m$, $s_k$ of length $j,m,k$, and are ordered as
\begin{equation}
s_A = s_j s_m \ ,  \quad \quad s_B = s_m s_k \ , \label{eq:s_As_B}
\end{equation}
i.e.\ such that the overlapping Majoranas are flushed to the right and left, respectively. This can be done without loss of generality, because this reordering merely costs a certain phase $\pm1$, and in the end the reordering can be undone, canceling the extra phase. Anticommutation relations imply
\begin{align}
s_As_B&=s_js_ms_ms_k = (-1)^{jm+mk+jk}s_ms_ks_js_m  \label{eq:overlapcommut1}\\
			&= (-1)^{(j+m)(k+m)-m^2}s_Bs_A	\ .		\label{eq:overlapcommut2}
\end{align}
The two strings commute if, and only if, $(j+m)(k+m)-m^2$ is even. This condition is equivalent to the length of the overlap having the same parity as the product of the string lengths, as in the main text.

\end{document}